\documentclass[openacc]{rstransa}%%%%where rstrans is the template name

\begin{document}

%%%% Article title to be placed here
\title{Searching for supersymmetry and its avatars}

\author{%%%% Author details
John~Ellis$^{1}$}%, X. Second author$^{2}$ and X. Third author$^{3}$}

%%%%%%%%% Insert author address here
\address{$^{1}$Theoretical Particle Physics and Cosmology Group, Department of
  Physics, King's~College~London, London WC2R 2LS, United Kingdom;\\
Theoretical Physics Department, CERN, CH-1211 Geneva 23,
  Switzerland;\\ 
  NICPB, R\"avala pst. 10, 10143 Tallinn, Estonia}

%%%% Subject entries to be placed here %%%%
%\subject{xxxxx, xxxxx, xxxx}

%%%% Keyword entries to be placed here %%%%
\keywords{Standard Model, monopole, supersymmetry}

%%%% Insert corresponding author and its email address}
\corres{John~Ellis\\
\email{John.Ellis@cern.ch}}

%%%% Abstract text to be placed here %%%%%%%%%%%%
\begin{abstract}
{\it Why continue looking for supersymmetry?} Over and above the aesthetic and theoretical
motivations from string theory, there are several longstanding phenomenological motivations for TeV-scale
supersy-mmetry such as the electroweak scale, and the lightest supersymmetric particle (LSP)
as cold dark matter. Run~1 of the LHC has actually provided three extra motivations, namely the stabilization of
the electroweak vacuum, and successful predictions for the Higgs mass and couplings. {\it How to look for it?}
There are several examples of emergent supersymmetry, the most recent being on the surfaces of topological 
insulators, and some sort of effective supersymmetry could be useful for boosting the power of laser arrays.
At the LHC, attention is moving towards signatures that had previously been neglected, such as long-lived
charged particles - which might be an opportunity for the MoEDAL experiment.
\end{abstract}
%%%%%%%%%%%%%%%%%%%%%%%%%%%

%%%%%%%%%% Insert the texts which can accomdate on firstpage in the tag "fmtext" %%%%%

%\begin{fmtext}

\maketitle

\section{Introduction}

In addition to its intrinsic elegance and role in string theory, there have been many phenomenological
arguments suggesting that supersymmetry might appear at an accessible energy scale, including its
ability to make the electroweak mass scale appear more natural~\cite{natural}, its provision of an interesting
dark matter candidate~\cite{EHNOS}, and its ability to facilitate the grand unification of particle interactions~\cite{GUT}.

However, in the absence (so far) of supersymmetry at the LHC, some physicists are questioning the primacy
of this paradigm for particle physics beyond the Standard Model (SM). Actually, I would argue the
opposite, namely that the discovery of the Higgs boson in Run~1 of the LHC has
provided three new motivations for supersymmetry at a potentially accessible energy scale. One new
motivation is vacuum stability. Feeding the Higgs mass measured at the LHC and the top quark mass
measured there and at Fermilab into renormalization-group calculations of the effective
potential in the SM suggest that our present electroweak vacuum is unstable~\cite{vacstab}, and raise the question
how the universe even arrived in this state~\cite{gothere}. These issues would be avoided in a supersymmetric
extension of the SM. Moreover, supersymmetry also predicted correctly the mass of the Higgs boson~\cite{SusyH} 
and that its couplings should resemble those in the SM~\cite{EHOW}.

However, before addressing the main subject of this paper, namely the search for supersymmetry at 
the level of fundamental particles, I also mention the appearance of supersymmetry in monopole physics
and a couple of avatars of supersymmetry at
less fundamental levels. One is an example of emergent supersymmetry in superconductivity, and the
other is an example of induced supersymmetry, which may have an application in laser technology.

\section{Supersymmetric Avatars}

There are many instances where supersymmetry emerges at an effective level, with examples including
nuclear and atomic physics~\cite{effectiveSUSY}. Supersymmetry also emerges in some theories with topological avatars.
One such example is provided by magnetic monopoles~\cite{BPS}:
there is a lower bound (BPS) on the monopole mass:
\begin{equation}
E \; \ge \; || \int_{S^2} {\rm Tr} [\phi {\bf B}.d \mathbf{S}] \, || \, ,
\label{BPS}
\end{equation}
which is saturated if the Higgs mass and potential vanish, as happens in $N=2$ supersymmetric theories.
BPS monopole solutions are generically supersymmetric, so maybe the MoEDAL experiment will discover
(at least approximate) supersymmetry at the same time as a monopole?

\begin{figure}[ht!]
\centering
\includegraphics[scale=0.4]{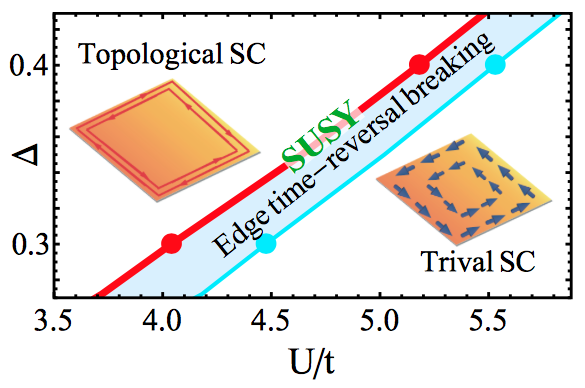}
\caption{\it Emergent supersymmetry in the quantum phase diagram of interacting topological superconductors. $N = 1$ space-time supersymmetry emerges at the boundary between the topological and trivial superconducting phases. Figure taken from~\cite{susySC1}.
}
\label{EmergentTopologicalSUSY}
\end{figure}

Another interesting recent example is provided by calculations~\cite{susySC1} and experiments~\cite{susySC2} 
that suggest the emergence of supersymmetry 
on the surfaces of topological insulators, at the boundary in parameter space between normal and topological
superconductors, as illustrated in Fig.~\ref{EmergentTopologicalSUSY}.

\begin{figure}[ht!]
\centering
\includegraphics[scale=0.6]{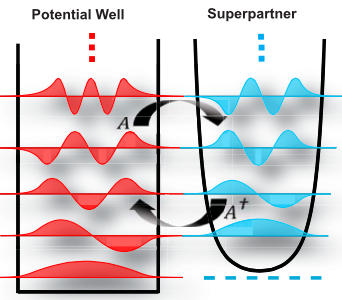}
\caption{\it Supersymmetric engineering: a semiconductor laser array is designed with coupled ``supersymmetric'' pairs of higher-energy modes,
lying above an unpaired fundamental mode. Figure adapted from~\cite{susylaser}.
}
\label{SUSYLaser}
\end{figure}

This is all very well, but what use is supersymmetry? Another recent paper has introduced the concept of
``supersymmetric engineering" with application to arrays of semiconductor lasers~\cite{susylaser}. The issue here is how
to concentrate the energy emission in a single mode. Inspired by supersymmetric quantum mechanics,
the proposed solution is to construct an array with identical spectra
at the levels of the $n > 1$, but with the lowest $n = 1$ mode unpaired, as
illustrated in Fig.~\ref{SUSYLaser}. In theory, all the energy should be
emitted in this lowest mode, and experiment indeed seems to show an enhancement in such a ``supersymmetric"
array~\cite{susylaser}. Maybe supersymmetry will turn out to be a useful idea, even before its discovery at a fundamental level?

\section{Searches for Fundamental Supersymmetry}

What about the searches for supersymmetry at the level of fundamental particle physics?
As is well known, there have been many experimental searches for supersymmetry at the LHC and elsewhere,
which have been unsuccessful so far~\cite{ATLASwiki,CMSwiki}. There have also been many searches for other possible extensions
of the SM, which have been equally fruitless. The supersymmetry searches have focused mainly on the
missing-energy signature that would be favoured if the lightest supersymmetric particle (LSP) provides the
astrophysical dark matter~\cite{EHNOS}.

This raises the questions whether it would be more productive to continue such missing-energy searches,
possibly looking more closely at some under-explored nooks of parameter space, or whether one should
focus on novel signatures?
One of the issues here is that there are many possible phenomenological manifestations of supersymmetry,
and there are no clear theoretical indications which to use as guidelines for experimental searches.
``There are no signposts in superspace."

The approach we have taken in the MasterCode Collaboration~\cite{MCweb} is to  compile all
the available experimental, phenomenological, experimental, astrophysical and cosmological constraints
that bear upon the possible masses of supersymmetric particles, and explore their implications in
frequentist statistical analyses of a range of different supersymmetric models. The relevant
measurements include include electroweak data, flavour observables, dark matter measurements including
the overall cosmological density of cold dark matter and upper limits on direct and indirect dark matter
searches, and the (so far) null results of LHC searches~\footnote{It has been suggested that weakly-interacting cold dark
matter of the type suggested by supersymmetry has issues with the absence of cusps and of satellite galaxies. However,
it has also been argued that there is in fact no cusp-core problem (see, e.g., \cite{Readcore}). nor any missing-satellite problem
(see, e.g., \cite{Readsatellite}). In the absence of consensus on these issues, here we stick with the supersymmetric cold dark matter paradigm.}.

\begin{figure}[ht!]
\centering
\includegraphics[scale=0.3]{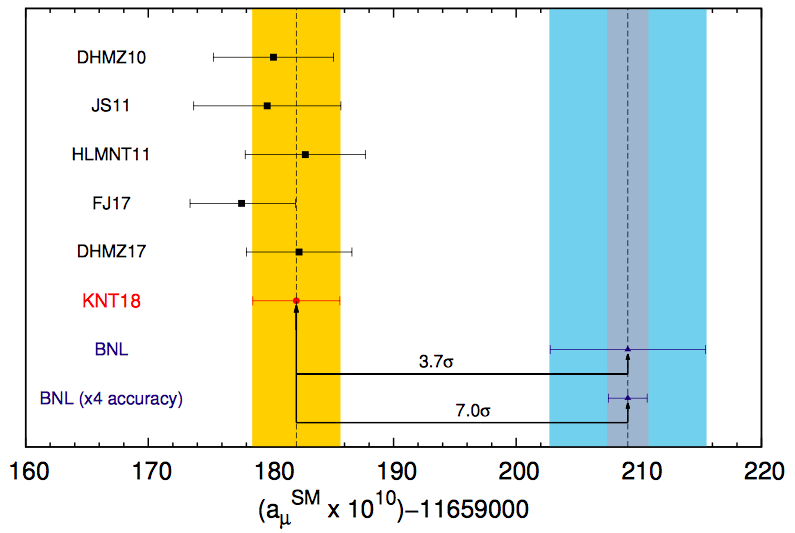}
\caption{\it Theoretical calculations of the anomalous magnetic moment of the muon, $g_\mu - 2 = 2 a_\mu$
in the Standard Model (yellow band on the left)~\cite{thgmu-2} disagree with the 
experiment measurement (blue band on the right)~\cite{expgmu-2}.
}
\label{gmu-2}
\end{figure}

Among all the laboratory measurements, there are none that provide unimpeachable evidence for new
physics beyond the SM. However, there are a couple of instances that merit closer attention. One is the
longstanding discrepancy between the experimental measurement~\cite{expgmu-2} and the SM calculation~\cite{thgmu-2}
of the anomalous magnetic moment of the muon, $g_\mu - 2$, illustrated in Fig.~\ref{gmu-2},
and another is the appearance of several 
anomalies in flavour physics. We look forward to experimental verification of the $g_\mu - 2$ discrepancy, 
which may soon be provided by an experiment at Fermilab~\cite{FNALgmu-2}. This discrepancy could be explained if there
are some electroweakly-interacting supersymmetric particles (sparticles) with low masses, but for the time being we treat it as an 
optional constraint on supersymmetric models. In parallel, we await clarification by the LHCb and Belle-2 
experiments  of the flavour anomalies, which would be difficult to explain in simple supersymmetric models.

\begin{figure}[ht!]
\centering
\includegraphics[scale=0.4]{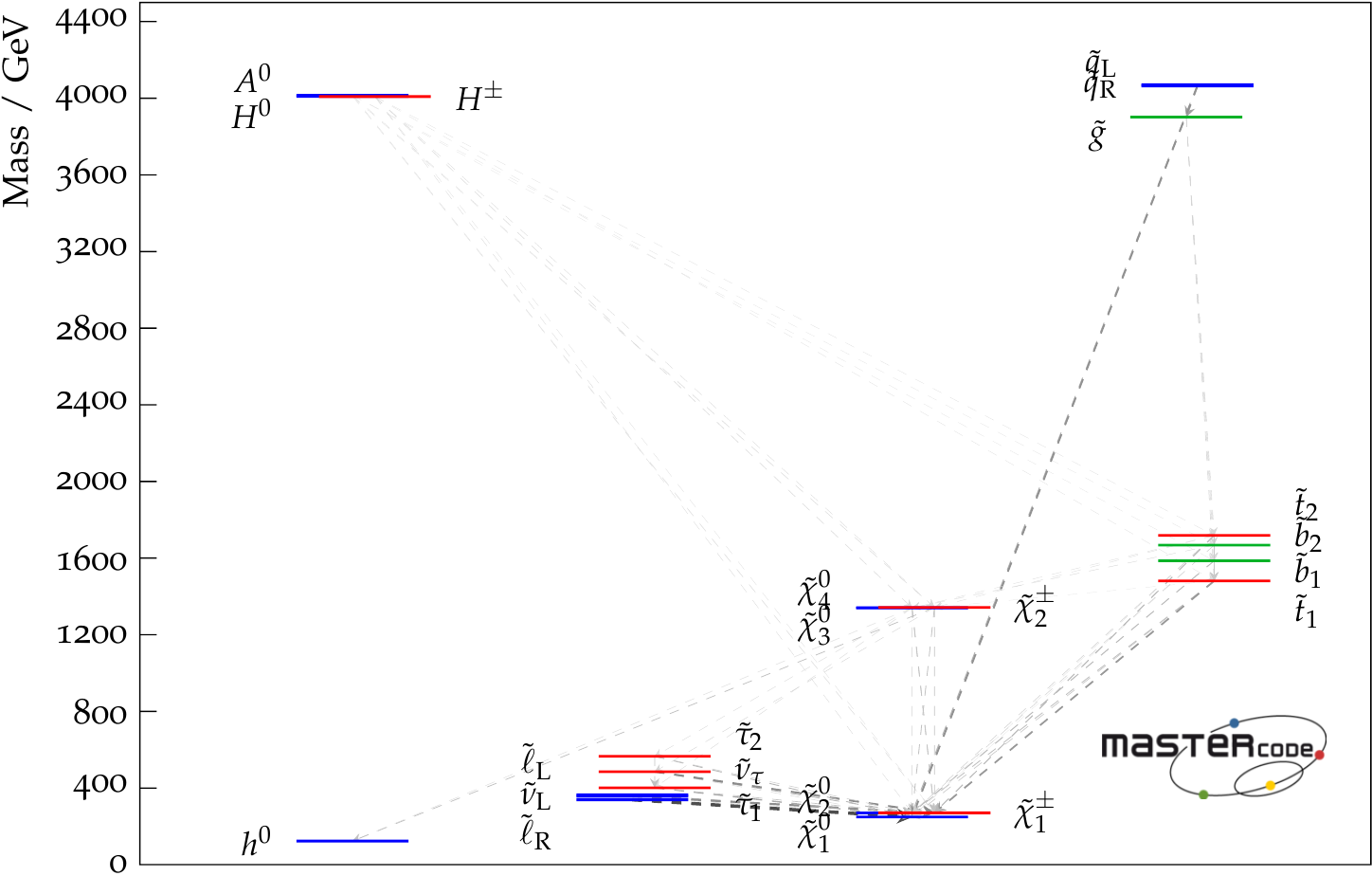}
\hspace{0.5cm}
\includegraphics[scale=0.4]{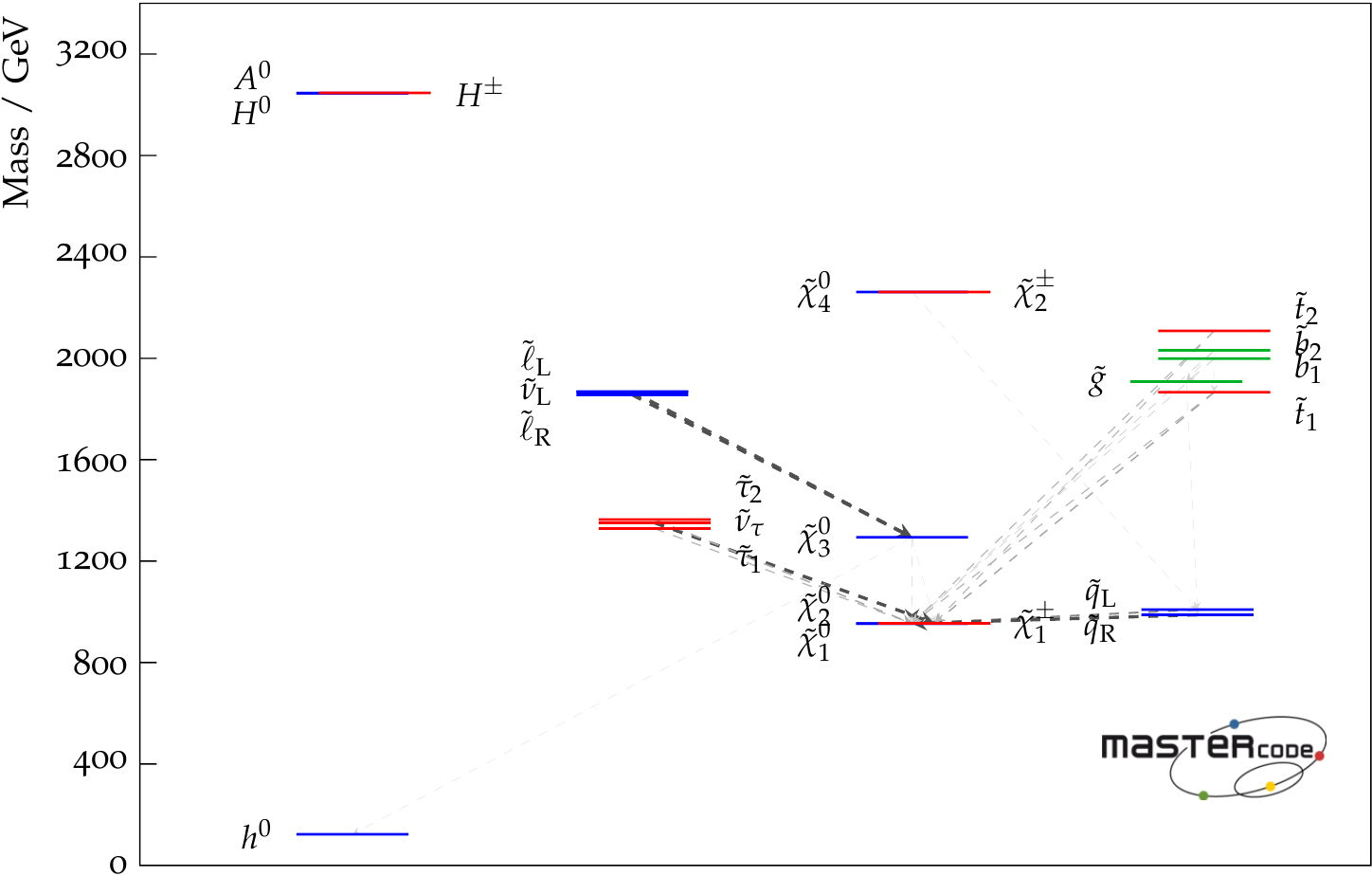}
\caption{\it Best-fit spectra in the pMSSM11 obtained in global fits including (dropping) the $g_\mu - 2$
constraint in the left (right) panel~\cite{MCpMSSM11}.
}
\label{spectra}
\end{figure}

One of the models we have studied is a phenomenological version of the minimal supersymmetric
extension of the SM with 11 parameters, the pMSSM11~\cite{MCpMSSM11}. We have analyzed its parameter space
with and without the $g_\mu - 2$ constraint, using a sample of $2 \times 10^9$ parameter sets.
Best-fit spectra in these two scenarios are shown in Fig.~\ref{spectra}~\cite{MCpMSSM11}.
Dropping $g_\mu - 2$, we found that several squarks could well have masses around 1~TeV, 
opening promising prospects for future LHC searches, as well charginos and neutralinos. If $g_\mu - 2$ is included
in the fit, sleptons, charginos and neutralinos could well have masses around 400~GeV, and these and some
squarks might be accessible to the LHC, whereas others might be out of its reach.
Our fits with and without $g_\mu - 2$ also offer some prospects for producing sparticles at the 3-TeV $e^+ e^-$
centre-of-mass energy proposed for CLIC, and the ILC operating at 1~TeV also has some prospects
in the fit with $g_\mu - 2$. However, the prospects for discovering supersymmetry at lower-energy $e^+ e^-$
colliders are not promising in either of our pMSSM11 analyses.

\begin{figure}[ht!]
\centering
\includegraphics[scale=0.33]{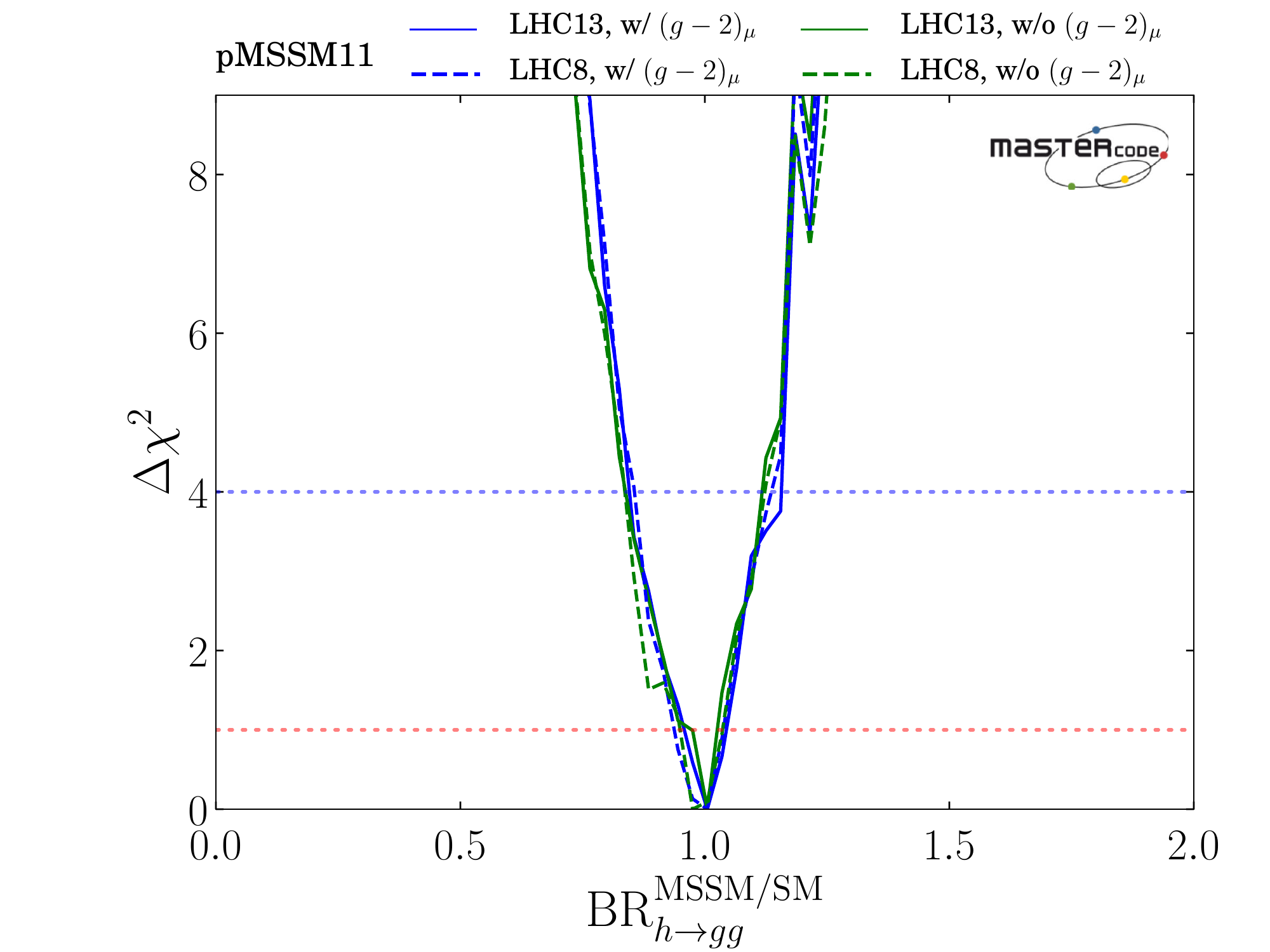}
\includegraphics[scale=0.33]{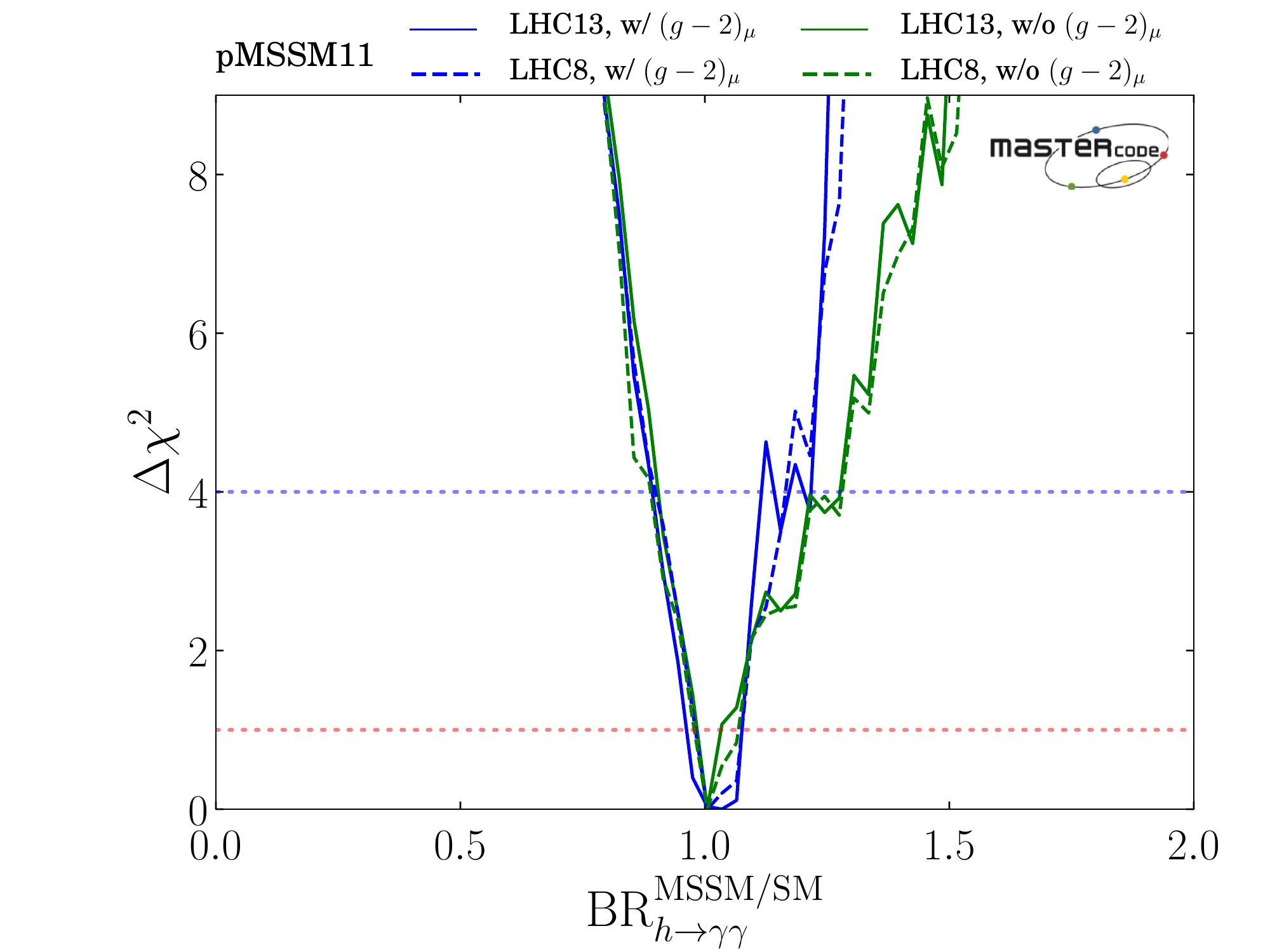}
\caption{\it One-dimensional $\chi^2$ likelihood functions for the branching ratios of the Higgs boson decays to
$gg$ (left panel) and $\gamma \gamma$ (rightt panel)  in global fits with/without the $g_\mu -2$ constraint (blue/green)
and including/dropping LHC Run~2 measurements (solid/dashed)~\cite{MCpMSSM11}.
}
\label{Higgs}
\end{figure}

There has been a lot of interest in the prospects for discovering the stop squark: arguments based on
the naturalness of the electroweak mass scale suggest that it might be relatively light, whereas some have argued
for a heavier stop squark on the basis on the Higgs mass measurement. We find that a stop squark weighing 
$\sim 1$~TeV is quite compatible with $m_H$ and other constraints at the $\Delta \chi^2 = 1$ (68\% confidence)
level, and that a range of masses around 500~GeV is allowed at the $\Delta \chi^2 = 4$ (95\% confidence)
level. We have also studied the likely ranges of Higgs decay branching ratios in our fits. As illustrated in
Fig.~\ref{Higgs}, in general we find
best-fit values that are very close to those in the SM, but 20\% deviations appear quite possible at the
$\Delta \chi^2 = 4$ (95\% confidence) level. These might be accessible to the upcoming higher-precision
Higgs measurements at HL-LHC.

\begin{figure}[ht!]
\centering
\includegraphics[scale=0.3]{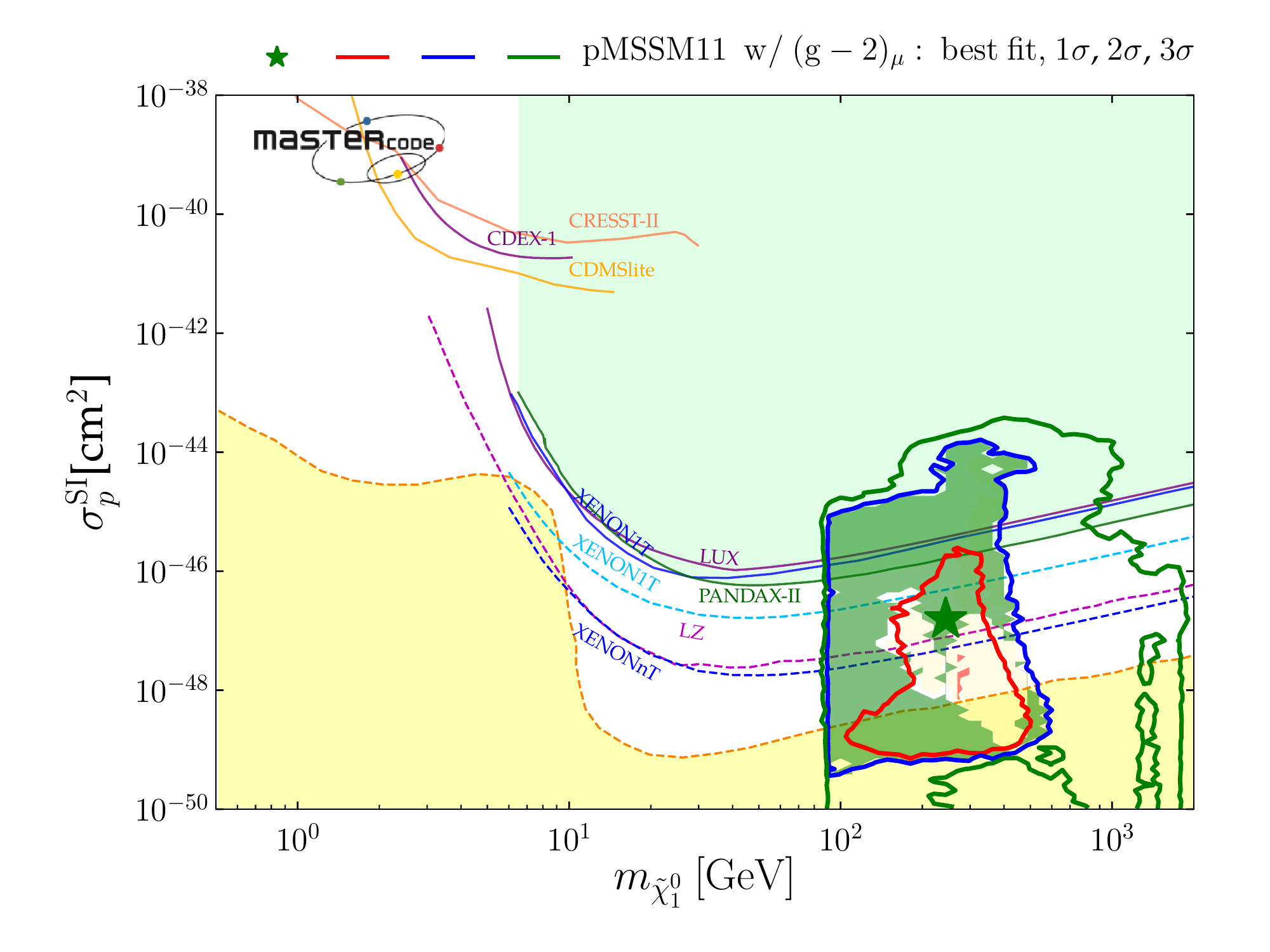}
\includegraphics[scale=0.3]{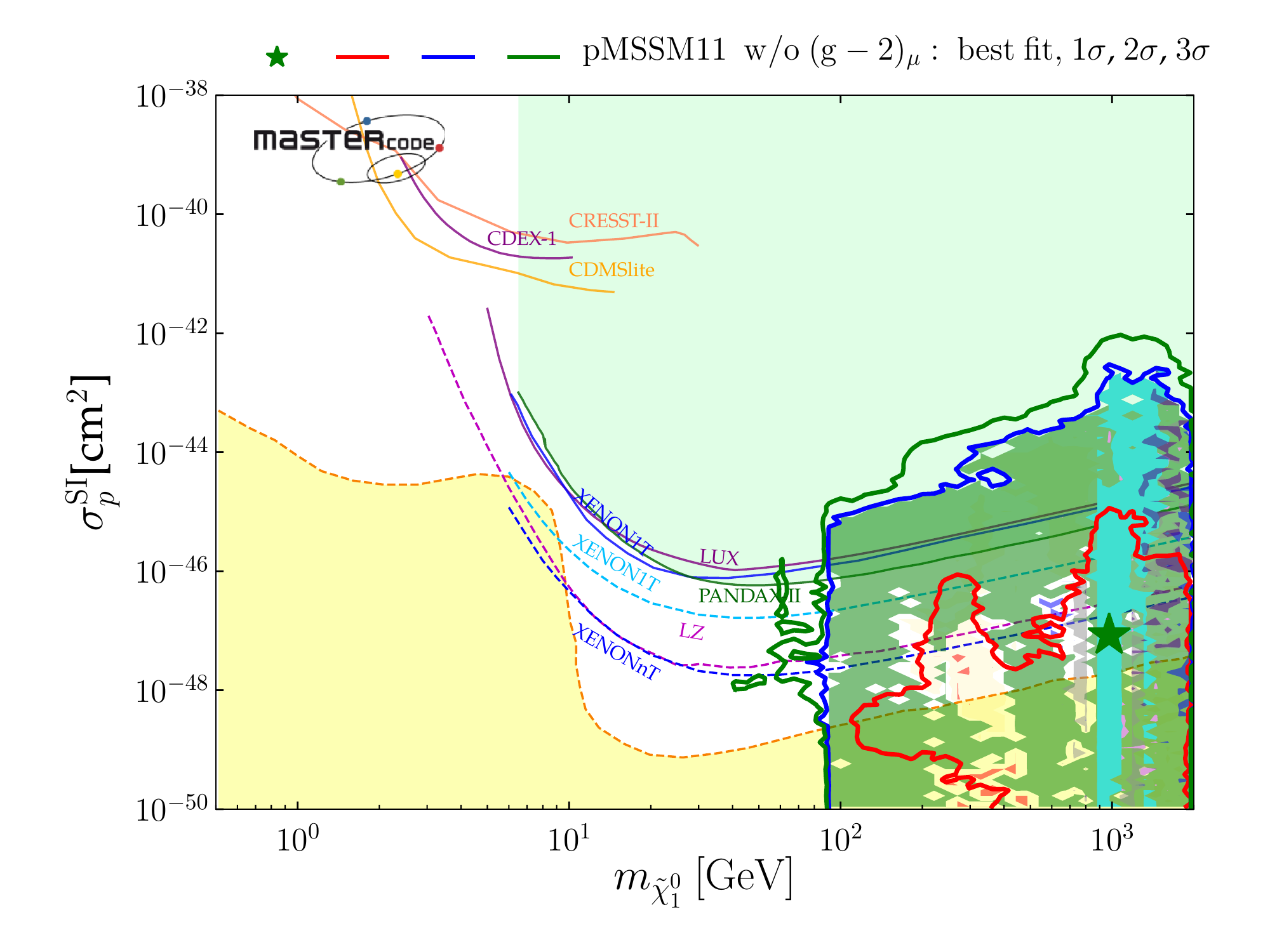}
\caption{\it Predictions for spin-independent dark matter scattering in global fits with (without) the $g_\mu - 2$
constraint in the left (right) panel, showing direct constraints (solid lines), the prospective sensitivities of future experiments
(dashed lines) and the neutrino `floor'~\cite{floor}. The 68\% (95\%) CL regions are outlined in red (blue),
and the colours inside these regions correspond to the dominant mechanisms controlling the cosmological
LSP density. See~\cite{MCpMSSM11} for details.
}
\label{ssi}
\end{figure}

Another important way to search for supersymmetry is to look directly for the scattering of LSP
dark matter particles on matter in a deep underground laboratory. The preferred mass range for the LSP
in our analysis is $\sim 300$~GeV if the $g_\mu - 2$ constraint is included, or $\sim 1$~TeV if is
dropped. Either way, the cross section for spin-independent dark matter scattering could be very close
to the present experimental upper limits~\cite{MCpMSSM11}, as seen in Fig.~\ref{ssi}. 
On the other hand, it might also be much smaller, below the
`floor'~\cite{floor} where astrophysical neutrino backgrounds become important.

How heavy could the LSP be? The cosmological density of dark matter is an important constraint,
which can be respected by a heavy LSP only if its rate of annihilation in the early universe is enhanced
in some way. This can happen if the LSP is nearly degenerate with the next-to-lightest supersymmetric
particle, the NLSP, and the two species coannihilate.  In such a case, the mass difference might be so
small that the  NLSP has a long lifetime for decay into the LSP.
There are other scenarios in which the NLSP might be long-lived, e.g., if the LSP is the gravitino in which
case the NLSP decay would be suppressed by a gravitation-strength coupling, or in split supersymmetric
scenarios in which the the sparticle mediating NLSP decay is very heavy. Alternatively, the LSP would itself
be unstable and long-lived if there is a small coupling violating $R$ parity. With all these motivations, there
has recently been increased interest in searches for long-lived unstable sparticles at the LHC.

\section{Anomalous Sparticle Signatures in the MoEDAL Experiment}

\begin{figure}[ht!]
\centering
\includegraphics[scale=0.35]{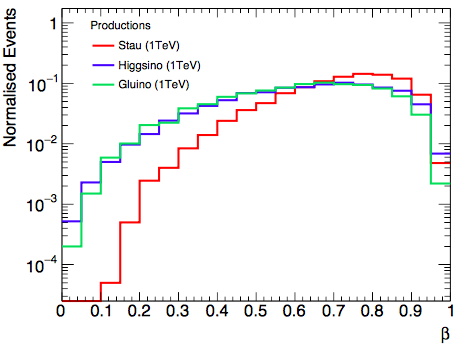}
\hspace{0.5cm}
\includegraphics[scale=0.3]{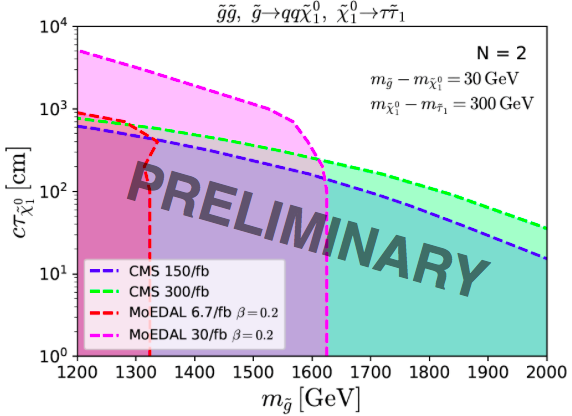}
\caption{\it Left panel: Calculations of the velocity distributions of sparticles at the LHC.
Right panel: Comparison between the sensitivities of MoEDAL and CMS to the production of
sparticles at the LHC. Figures taken from~\cite{KS}.
}
\label{MoEDAL}
\end{figure}

Whilst MoEDAL has been designed to optimize its ability to detect magnetic monopoles, it also has
capabilities to detect other heavily-ionizing particles~\cite{MoEDAL}. In particular, MoEDAL's nuclear track detectors 
(NTDs) are sensitive to the relatively high ionization from slow-moving singly-charged particles.
with velocities $\beta < 0.2$. The stau slepton,
${\tilde \tau}$, is a prime candidate to be the NLSP.  Unfortunately most directly-produced ${\tilde \tau}$s
would be produced with larger values of $\beta$, as seen in the left panel of Fig.~\ref{MoEDAL}~\cite{KS}. 
Therefore, MoEDAL has relatively low efficiency $\epsilon$
and hence sensitivity to direct ${\tilde \tau}$ production:
\begin{equation}
\epsilon \cdot \sigma \; \sim \; (< 10^{-3}) \cdot (< 100)/{\rm fb}
\end{equation}
for $m_{\tilde \tau} > 100$~GeV. However, the picture improves for ${\tilde \tau}$s that
are produced indirectly via the cascade decays of heavier sparticles, 
as illustrated in the left panel of Fig.~\ref{MoEDAL}~\cite{KS} - which is actually
expected to be the dominant production mechanism. As a result, at the end of Run~3
of the LHC, when ATLAS and CMS hope to have gathered $\sim 300$/fb, MoEDAL may
have comparable sensitivity to some supersymmetric scenarios with long-lived ${\tilde \tau}$s,
as seen in the right panel of Fig.~\ref{MoEDAL}~\cite{KS}.

MoEDAL is also installing a complementary detector for penetrating particles,
MoEDAL  Apparatus for Penetrating Particles (MAPP)~\cite{MAPP}. This will search for
long-lived neutral particles, particles with electric charges $\ll e$, and other
anomalously-penetrating particles. A demonstration detector was installed in December 2017,
and the full detector will be ready for Run~3 of the LHC. This is one of a number of approved~\cite{FASER}
and proposed~\cite{others} experiments at the LHC to look for long-lived, weakly-interacting particles~\cite{LHC-LLP},
which can probe supersymmetric scenarios in which the NLSP is almost degenerate with the LSP,
is neutral and has no strong interactions 

\section{Longer-Term Prospects for Supersymmetry}

The LHC will continue to run into the mid-2030s, aiming to accumulate in ATLAS and
CMS $\gtrsim 20$ times more than the $\sim 140$/fb that they have accumulated so far, not all of
which has been analyzed for many supersymmetric signatures. Theoretically, it is
certainly possible that sparticles may be lying beyond the current reaches of ATLAS
and CMS, but within reach of future LHC runs. This could happen, for example, if the
NLSP is a stop squark that is almost degenerate with the LSP. MoEDAL will continue
its parallel searches for particles with anomalous ionization signatures.

\begin{figure}[ht!]
\centering
\includegraphics[scale=0.5]{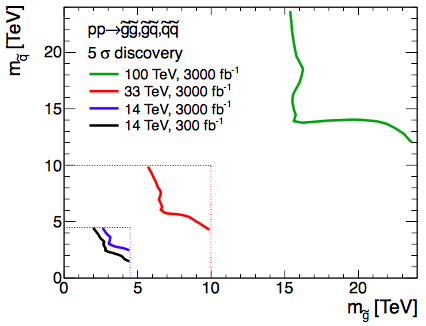}
\caption{\it Estimated 5-$\sigma$ discovery reaches for squarks and gluinos at the LHC (14 TeV),
HE-LHC (33 TeV) and FCC-hh (100 TeV)~\cite{FCC-hh1}. 
}
\label{squarkgluino}
\end{figure}

There are many ongoing discussions about possible high-energy colliders beyond the LHC.
One possibility being discussed actively at CERN is a large circular tunnel able to accommodate
a collider for electrons and positrons at relatively low energies but very high luminosities (FCC-ee)~\cite{FCC-ee},
and/or  a collider for protons at 100~TeV in the centre of mass (FCC-hh)~\cite{FCC-hh1,FCC-hh}, also with a very high luminosity. These
 will enable the search for supersymmetry to be carried into the range above 10~TeV, via
 both direct searches~\cite{FCC-hh1,FCC-hh} (see Fig.~\ref{squarkgluino}) and indirect probes~\cite{FCC-ee}.
 
%\end{fmtext}

\subsection*{Acknowledgements}

\noindent
The work of JE was supported in part by the United Kingdom STFC Grant
ST/P000258/1, and in part by the Estonian Research Council via a
Mobilitas Pluss grant.

\end{document}